\documentclass[a4paper,prl,twocolumn,longbibliography,superscriptaddress]{revtex4-1}
\usepackage{amsmath}
\usepackage{amssymb}
\usepackage{graphicx}
\usepackage{bm}
\usepackage{natbib}
 \usepackage[normalem]{ulem}

\usepackage{hyperref}

\hypersetup{
    colorlinks=true,
    linkcolor=blue,
    filecolor=magenta,      
    urlcolor=cyan,
}

\usepackage[usenames,dvipsnames]{xcolor}

 \newcommand{\beq}{\begin{equation}}
  \newcommand{\eeq}{\end{equation}}
 \newcommand{\pcollapse}{p_{\mathrm{collapse}}}

 \newcommand{\Pload}{{\cal F}}

 \newcommand{\nmax}{n_{\mathrm{max}}}
 \renewcommand{\t}{h}
  
 \newcommand{\tpinfty}{\tilde{\Pload}_{0}}

    \newcommand{\bt}{\tilde{t}}
    \newcommand{\btfreeze}{\bt_{\mathrm{form}}}

 \newcommand{\pring}{\tilde{p}_{\mathrm{ring}}}
 \newcommand{\pann}{\tilde{p}_{\mathrm{ann}}}
 \newcommand{\Tin}{T_{\mathrm{in}}}
  \newcommand{\Tout}{T_{\mathrm{out}}}
 \newcommand{\Rout}{R_{\mathrm{out}}}
 \newcommand{\Rin}{R_{\mathrm{in}}}
  \newcommand{\kmax}{k_{\ast}}
  \newcommand{\smax}{\sigma_{\mathrm{max}}}
 \newcommand{\srr}{\Sigma_{rr}}
  \newcommand{\sqq}{\Sigma_{\theta\theta}}

\begin{document}

\title{Dynamic buckling of an elastic ring in a soap film}

\author{Finn Box}
\affiliation{Mathematical Institute, University of Oxford, Woodstock Rd, Oxford, OX2 6GG, UK}
\author{Ousmane Kodio}
\affiliation{Mathematical Institute, University of Oxford, Woodstock Rd, Oxford, OX2 6GG, UK}
\affiliation{Department of Mathematics, Massachusetts Institute of Technology, Cambridge, Massachusetts 02139, USA}
\author{Doireann O'Kiely}
\affiliation{Mathematical Institute, University of Oxford, Woodstock Rd, Oxford, OX2 6GG, UK}
\affiliation{Department of Mathematics and Statistics, University of Limerick, Limerick, V94 T9PX, Ireland}
\author{ Vincent Cantelli}\affiliation{Mathematical Institute, University of Oxford, Woodstock Rd, Oxford, OX2 6GG, UK}
\author{ Alain Goriely}
\affiliation{Mathematical Institute, University of Oxford, Woodstock Rd, Oxford, OX2 6GG, UK}
\author{Dominic Vella}\email[]{dominic.vella@maths.ox.ac.uk}
\affiliation{Mathematical Institute, University of Oxford, Woodstock Rd, Oxford, OX2 6GG, UK}

\begin{abstract}
 Dynamic buckling may occur when a load is rapidly applied to, or removed from, an elastic object at rest. In contrast to its static counterpart, dynamic buckling offers a wide range of accessible patterns depending on the parameters of the system and the dynamics of the load. To study these effects, we consider experimentally the dynamics of an elastic ring in a soap film when part of the film is suddenly removed. The resulting change in tension applied to the ring creates a range of interesting patterns that cannot be easily accessed in static experiments. Depending on the aspect ratio of the ring's cross section, high-mode buckling patterns are found in  the plane of the remaining soap film or out of the plane.  Paradoxically, while inertia is required to observe these non-trivial modes, the selected pattern  does not depend on inertia itself. The evolution of this pattern beyond the initial instability is studied experimentally and explained through theoretical arguments linking dynamics to pattern selection and mode growth. We also explore the influence of dynamic loading and show numerically that by imposing a rate of loading that competes with the growth rate of instability, the observed pattern can be selected and controlled.
\end{abstract}

\maketitle

\paragraph{Introduction} A striking feature of  elastic buckling  is the appearance of wrinkle patterns with a well-defined wavelength; these regular patterns have found a range of applications in the design of structures across length scales \cite{Reis2015,Kim2012}, as well as in the measurement of material properties \cite{Stafford2004,Huang2007}. While compression is required to induce buckling, the wavelength of static wrinkle patterns  emerges from a trade-off between an object's resistance to bending and a resistance to out-of-plane displacement \cite{Cerda2003}. The resulting balance governing pattern formation is therefore usually established by the material properties of the system alone, with few means to change the pattern formed without changing material properties. As a concrete example, consider the compression of a stiff, thin sheet attached to a soft substrate; the wavelength of the resulting wrinkles is set by the ratio of the two elastic moduli, together with the thickness of the thin sheet \cite{Stafford2004} --- while a compressive force is required to cause buckling, the wrinkling wavelength varies only weakly with the applied load in such static scenarios \cite{Davidovitch2011}.

\begin{figure}[h]
	\centering
  \includegraphics[width=0.85\linewidth]{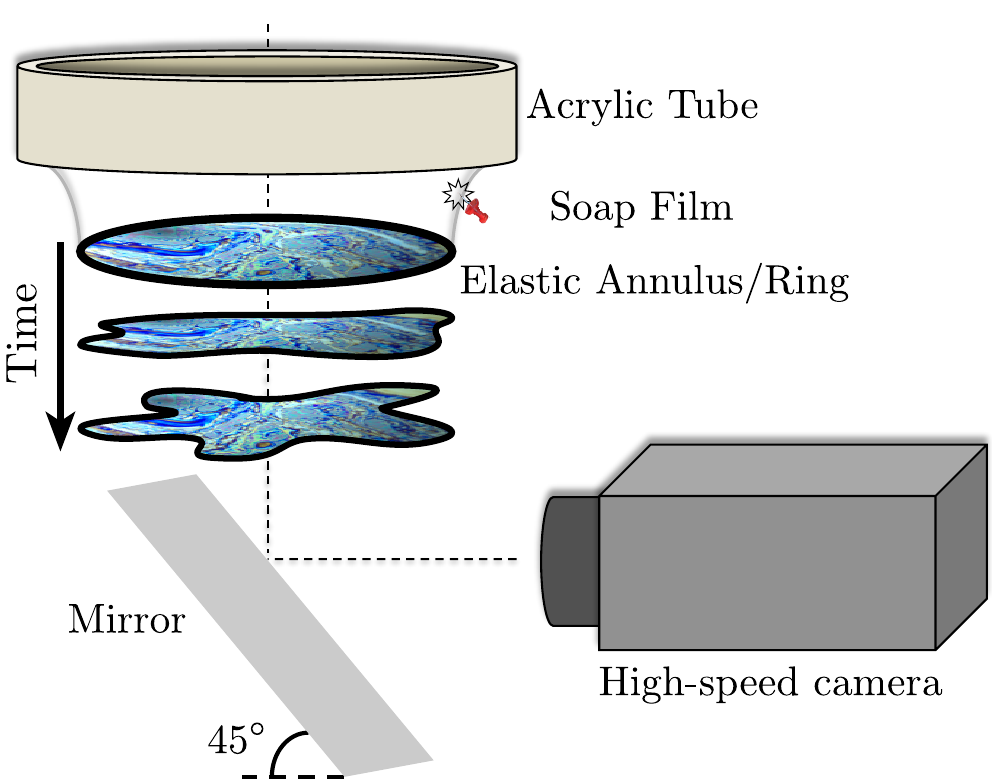}
	\caption{General principle of the experiment: a closed elastic object (a ring or annulus) is held within a soap film. At $t=0$, the outer soap film is broken allowing the object, with inner soap film intact, to fall freely.  The unbalanced surface tension force from the inner soap film causes the object to buckle dynamically; this motion is captured by a high speed camera.}
	\label{fig:setup}
\end{figure}

\begin{figure*}
  \centering
  \includegraphics[width=0.8\linewidth]{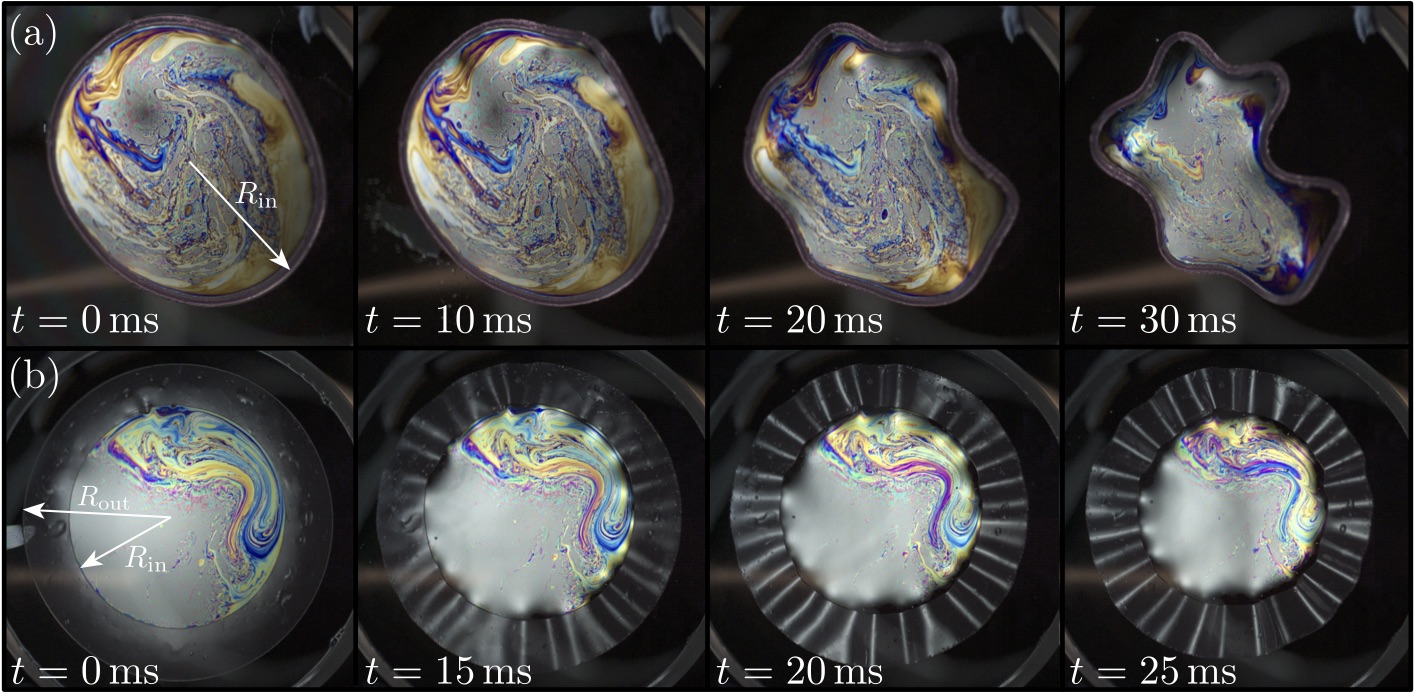}
  \caption{Experimental snap-shots  from the dynamic buckling of an elastic (a) ring ($\t=1\mathrm{~mm}=\Rout-\Rin$, $\Rin=23.5\mathrm{~mm}$, $E=42\mathrm{~kPa}$) and (b) annulus ($\t=110\mathrm{~\mu m}$, $\Rin=25\mathrm{~mm}$, $\Rout=34\mathrm{~mm}$, $E=200\mathrm{~kPa}$) under the action of an unbalanced surface tension $\Tin$. Images are shown at various intervals of time after the bursting of the outer soap film \cite{Note2}. In (a) the dimensionless pressure $\pring=\Tin\Rin^3/(B\t)=98$, while in (b) the dimensionless pressure $\pann=\Tin\Rin^2/B=528$ and $\Rout/\Rin=1.36$, with $B$ the bending stiffness; our linear stability analysis predicts a most unstable mode number $n=7$ and $n=30$ in (a) and (b), respectively. }
  \label{fig:pics}
\end{figure*}

Dynamic buckling is, however, known to give an alternative route to pattern selection \cite{Lindberg1987,Gladden2005,Box2013}. For example, when stretched and released, a rubber band buckles with a well-defined wavelength  \cite{Vermorel2007,Oratis2019}, as does a rubber sheet subject to a normal impact \cite{Vermorel2009}, while impacting a brittle rod at one end leads to the formation of fragments with well-defined sizes \cite{Gladden2005}. In each of these cases there is no resisting substrate, but the compressive force within the object, $\Pload$, plays a similar role. However, in each of these scenarios, $\Pload$ is not directly controlled and must be determined; moreover, $\Pload$ evolves along with buckling \cite{Gladden2005,Vermorel2009,Putelat2014}, further complicating the problem. In this Letter, we present an experimental system that allows a more direct probe of dynamic buckling. 

\paragraph{Experiment} Our experimental system consists of an elastic ring of inner and outer radii $\Rin$ and $\Rout$, respectively,  thickness $\t$ and Young's modulus $E$ placed within a soap film of surface tension $\gamma$ (fig.~\ref{fig:setup}). With soap film filling the inner, $r<\Rin$, and outer, $r>\Rout$, voids, the ring is in equilibrium, with a uniform and isotropic tension $\Tin=\Tout=\gamma$ throughout. However, if the outer soap film is broken (keeping the inner soap film intact) \footnote{Once the outer film is broken, the whole system is in free-fall for $\gtrsim0.1\mathrm{~s}$, which is longer than the times considered here.} then $\Tout=0$, with $\Tin>0$  unbalanced, pulling the ring  inwards. With stiffer boundaries, globally non-flat static solutions are known to exist \cite{Giomi2012,Chen2014}; instead of these, we observe relatively high mode number  buckling, $n>2$. Snap-shots of buckling (fig.~\ref{fig:pics} and see also Supplemental Material \footnote{Movies of buckling are available as Supplementary Information, together with details of the derivation of \eqref{eqn:annulus} and illustrative numerical results with ramped loading \eqref{eqn:RampProfile}.}) illustrate that the form of instability depends on the geometry of the ring: for an elastic ring of square cross-section (i.e.~$\Rout-\Rin=\t$), instability occurs via in-plane buckling, while for a wide, but thin, ring  with $\Rout-\Rin\gg \t$ (termed an annulus hereafter), instability occurs as out-of-plane  wrinkling. Here we report the results of two sets of experiments: (i) with elastic rings, of square cross-section $0.5\mathrm{~mm}\leq\t\leq1\mathrm{~mm}$ and inner radius $17.5\mathrm{~mm}\leq\Rin\leq32\mathrm{~mm}$ and (ii) with elastic annuli of thickness $43\mathrm{~\mu m}\leq\t\leq210\mathrm{~\mu m}$  and inner and outer radii $3\mathrm{~mm}\leq\Rin\leq25\mathrm{~mm}$ and $9\mathrm{~mm}\leq\Rout\leq 35\mathrm{~mm}$, respectively. These elastic structures were fabricated by casting and spin-coating Polyvinyl Siloxane (Elite Double 8, 22 \& 32, Zhermack, with Young's modulus $37\mathrm{~kPa}\leq E\leq800\mathrm{~kPa}$, measured via indentation, and Poisson ratio $\nu=0.5$); the surface tension of the soap film was measured, using a Wilhelmy plate, to be $\gamma=26.5\pm2.5\mathrm{~mN\,m^{-1}}$.

The patterns observed dynamically are qualitatively similar to those observed in the analogous static systems: for  a thin-walled elastic  cylinder of infinite length (or a ring of rectangular cross section) subject to an externally applied pressure it is known  \cite{Windenburg1934,Lindberg1987,Amabili2003} that the pressure at which buckling occurs is  $\pcollapse=3B/\Rin^3$ with  $B=E\t^3/12$  its bending stiffness. Just above this collapse pressure, the static system adopts a figure-of-eight shape, ultimately leading to self-contact \cite{Carrier1947,Flaherty1972}. While other higher modes of instability exist, they have only been observed by the imposition of an additional breaking of symmetry, e.g.~via confinement with a polygonal boundary of the desired mode number \cite{Hazel2017}. For thin, unconfined elastic annuli, static wrinkling patterns with higher modes of instability are common \cite{Cerda2003,Huang2007,Pineirua2013,Paulsen2017}, but require two opposing (though not in general equal) tensions to be applied: when $\Tout/\Tin\ll1$ in these static systems, the annulus breaks its gross axisymmetry by forming a lenticular or `stadium' shape \cite{Paulsen2017}.  Since the corresponding static scenarios lead to instability with $n=2$, the observation in fig.~\ref{fig:pics} of patterns with mode number $n>2$ is suggestive of a dynamic phenomenon.

\paragraph{Theory and Results} The essential mechanism of dynamic buckling can be understood by considering a one-dimensional beam of bending stiffness $B$ and  density $\rho$ that is aligned along the $x$-axis \cite{Lindberg1987}. Small beam deflections, $\zeta(x,t)$, caused by a constant imposed compressive force per unit length, $\Pload$, satsify \cite{Howell2009}
\beq
\rho \t\frac{\partial^2\zeta}{\partial t^2}=-B\frac{\partial^4\zeta}{\partial x^4}-\Pload \frac{\partial^2\zeta}{\partial x^2}.
\label{eqn:Beam}
\eeq Neglecting boundary conditions \cite{Lindberg1987}, one finds that the growth rate $\sigma$ of a perturbation with wave number $k$ satisfies the dispersion relation
\beq
\rho \t\sigma^2=-Bk^4+\Pload k^2.
\label{eqn:GrowthRate}
\eeq From this dispersion relation, it is  clear that all modes with wavenumber $0<k<(\Pload/B)^{1/2}$ are linearly unstable and, further, that the fastest growing mode has $k=\kmax=[\Pload/(2B)]^{1/2}$ with growth rate $\smax=[\Pload^2/(4B\rho \t)]^{1/2}=\kmax^2(B/\rho \t)^{1/2}$.


\begin{figure}
  \centering
  \includegraphics[width=0.8\columnwidth]{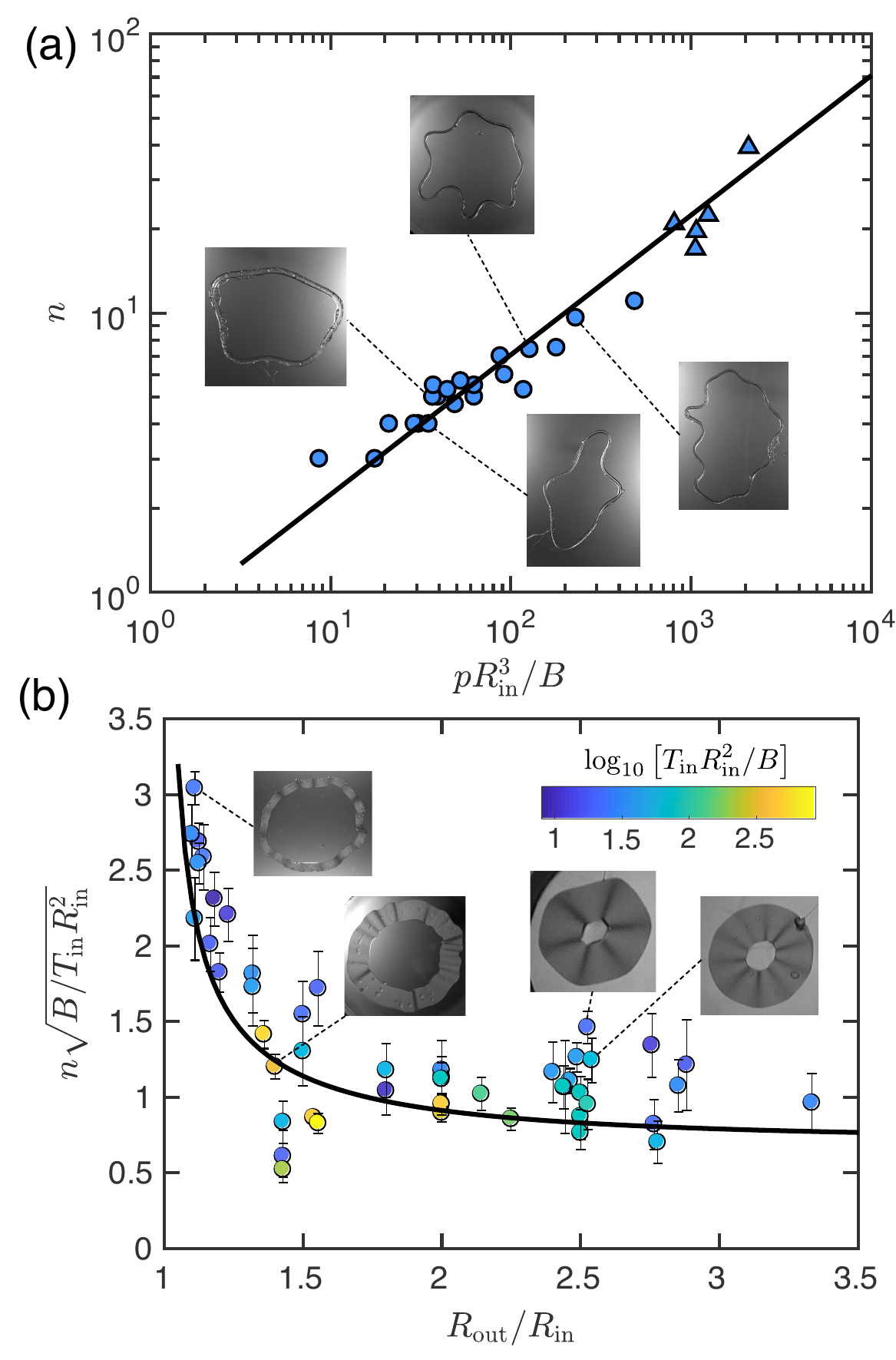}
  \caption{Experimentally observed mode numbers for dynamic buckling of a ring within a soap film. (a) Experimental results  for an elastic ring (points), together with the  prediction of the simplified linear stability analysis \eqref{eqn:Nring} (line). (Here, circles denote experiments for which $n$ is measured around the whole ring; triangles denote experiments for which $n$ is inferred from the wavelength of instability at onset \cite{Note3}.) (b) Experimental results  for an elastic annulus (points), show that the rescaled mode number $n\pann^{-1/2}$ depends on the width of the annulus $\Rout/\Rin$, as predicted by \eqref{eqn:annulus} (solid curve). (Here data points are colored by the value of $\pann=\Tin\Rin^2/B$, as in the inset colorbar; $\pann$ is used to rescale  $n$.) }
  \label{fig:Results}
\end{figure}

Variants of the above mechanism have been studied previously including: over-damped motion with a spatially-varying compression \cite{Kantsler2012} as well as the crucial role of imperfections \cite{Hutchinson1966} and plastic deformation \cite{Lindberg1987,Karagiozova2008} in many engineering applications. For the dynamic buckling of a ring, previous work has established the unstable mode numbers \cite{Wah1970} but here, we focus on understanding the mode number  observed, as well as the evolution beyond the onset of instability while the system remains elastic. To relate our ring experiment to the dynamic buckling of a rod just discussed, we note that the tension applied by the inner soap film, $\Tin=\gamma$, behaves as a thickness-averaged pressure $p=\Tin/\t$, which, in combination with the ring curvature, induces an in-plane compression $\Pload=p\Rin$ (the analogue of Laplace's law \cite{Boys1958}) within the ring. Neglecting other effects of the ring curvature, we expect that the observed wavelength of instability $\lambda=2\pi/\kmax=2^{3/2}\pi\left[B/(p\Rin)\right]^{1/2}$, or, equivalently, the number of buckles
\beq
n=\Rin\kmax=2^{-1/2}\left(\frac{\Tin\Rin^3}{B\t}\right)^{1/2}=\left(\frac{\pring}{2}\right)^{1/2}
\label{eqn:Nring}
\eeq where $\pring=\Tin\Rin^3/(B\t)$ is a dimensionless measure of the strength of the soap film's surface tension to the ring's bending stiffness \cite{Giomi2012,Chen2014}. This result is in reasonable agreement with the experimental results shown in fig.~\ref{fig:Results}a \footnote{For large $\pring$, coarsening of the buckles in one portion of the ring occurred before the outer soap film had fully retracted. In such cases, $n$ was inferred from the measured buckling wavelength at onset; in other cases, the value of $n$ reported is the maximum number of buckles observed.} and is distinct from the $\lambda\sim\t$ scaling observed in rubber band recoil \cite{Vermorel2007}. A more detailed model, incorporating the full effects of the ring curvature \cite{Kodio2019}, gives $\nmax$ indistinguishable from the  simplified result \eqref{eqn:Nring} for $n\gtrsim2$.

The situation for an annulus ($\Rout-\Rin\gg \t$) is  more involved since the stress varies spatially. While wrinkles remain small, we expect the state of stress to remain close to that of the planar Lam\'{e} problem \cite{Paulsen2017}; in particular the radial and hoop stresses
\beq
\Sigma_{rr,\theta\theta}=-\frac{\Tin}{\Rout^2/\Rin^2-1}\left(1\mp\frac{\Rout^2}{r^2}\right).
\eeq Note that the radial stress is tensile, $\srr>0$, and the hoop stress compressive $\sqq<0$, throughout the annulus $\Rin<r<\Rout$. Applying the result $\kmax=(\Pload/2B)^{1/2}$ locally with $\Pload=-\Sigma_{\theta\theta}(r)$ gives an expected number of wrinkles at the inner edge \cite{Note2}
\beq
\label{eqn:annulus}
n(\Rin)=\left(\frac{\Tin\Rin^2}{B}\right)^{1/2}\left[\frac{\Rout^2/\Rin^2+1}{2(\Rout^2/\Rin^2-1)}\right]^{1/2}
.
\eeq The result in \eqref{eqn:annulus} may be written as $n=\pann^{1/2}f(\Rout/\Rin)$ with the appropriate dimensionless pressure $\pann=\Tin\Rin^2/B$ and the effect of the width of the annulus characterized by the dimensionless function $f(x)=[(x^2+1)/2(x^2-1)]^{1/2}$. Experimental results with $8\lesssim\pann\lesssim900$ show a good collapse when $n\pann^{-1/2}$ is plotted as a function of $\Rout/\Rin$ (fig.~\ref{fig:Results}b); moreover, the dependence on $\Rout/\Rin$ is close to that expected from \eqref{eqn:annulus} (solid curve). Interestingly, as  $\Rout-\Rin\to \t$ the expression in \eqref{eqn:annulus} converges to  \eqref{eqn:Nring}, even though the modes of instability are different (in-plane buckling versus out-of-plane wrinkling).

Crucially, the results of \eqref{eqn:Nring} and \eqref{eqn:annulus} are very different from the corresponding quasi-static situations in which two-fold symmetry has been reported \cite{Flaherty1972,Paulsen2017}. The key ingredient that distinguishes static scenarios from those considered here is the presence of the elastic object's inertia. Given this importance of inertia in generating an instability qualitatively different from the static scenario, it seems paradoxical that the inertia itself does not appear to play a role in the final mode selection (eqns \eqref{eqn:Nring} and \eqref{eqn:annulus} are independent of the object's density $\rho$). To understand this paradox, we note from \eqref{eqn:GrowthRate} that the growth rate of the fastest growing mode $\smax=\sigma(\kmax)=\kmax^2\left[B/(\rho \t)\right]^{1/2}$,  \emph{does} depend on inertia. Our experiments and theoretical arguments assume that the unbalanced tensile load is applied very fast compared to the associated time scale of instability. We therefore hypothesize that imposing the applied load at different rates might give rise to additional control over the observed instability mode, but begin by first measuring the growth rate of instability experimentally.

For an elastic ring the growth  rate of the most unstable mode is
\beq
\smax =\left(\frac{B}{\rho \t\Rin^4}\right)^{1/2} \times\nmax^2\approx\left(\frac{B}{\rho \t\Rin^4}\right)^{1/2} \frac{\pring}{2},
\label{eqn:Sigma}
\eeq when $\pring\gg1$: for fixed material properties, the growth rate of instability increases with dimensionless pressure.  To measure the growth rate of instability, $\sigma$, experimentally, we monitor the area of the central hole enclosed by the ring, $A(t)$, which decreases during the experiment. A weakly-nonlinear analysis \cite{Kodio2019} shows that the change in this area from its initial value, $A_0=\pi\Rin^2$, initially grows at twice the growth rate of instability, i.e.~that
\beq
\frac{A_0-A(t)}{A_0}\sim \exp(2\smax t)
\label{eqn:Area}
\eeq for $\smax t\ll1$.

\begin{figure}[h]
  \centering
   \includegraphics[width=0.9\linewidth]{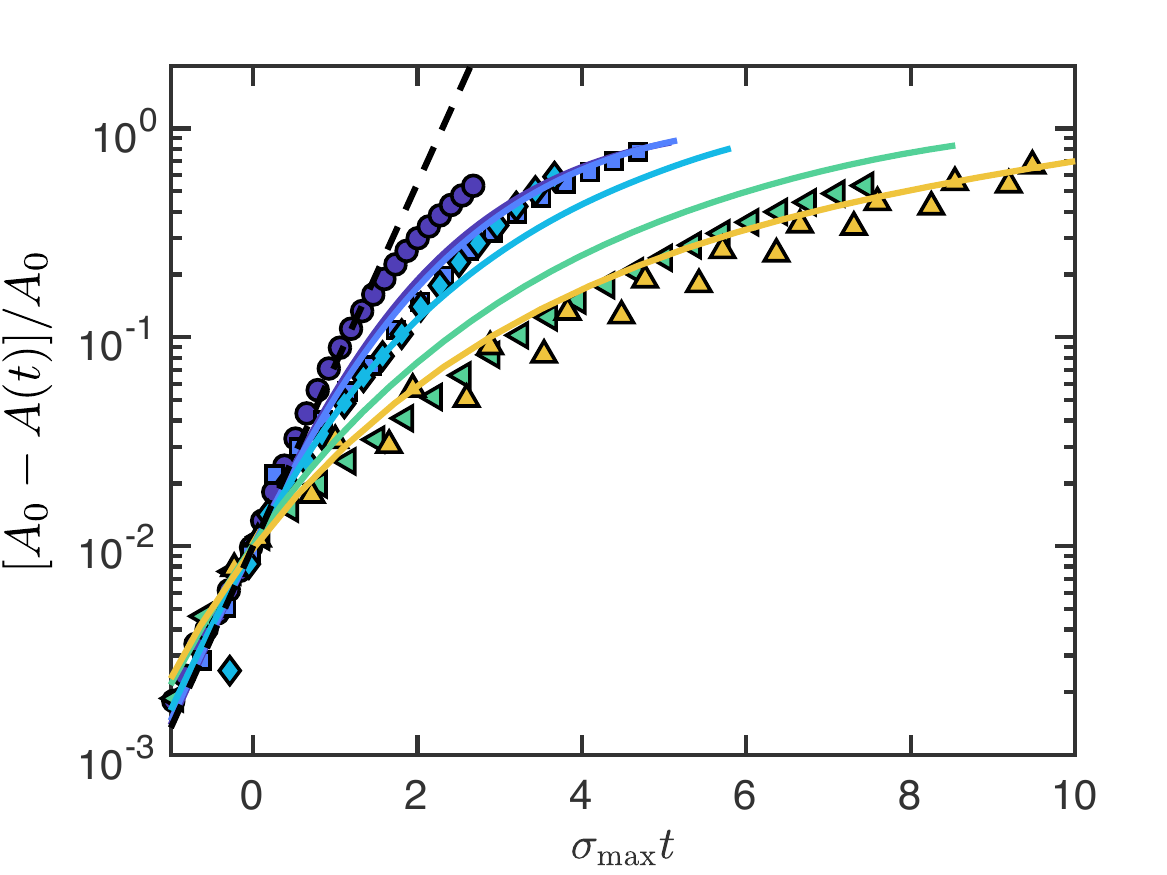}
        \caption{The growth rate of instability in an elastic ring is observed via the relative change of area, $\Delta A/A_0$, as
 a function of the rescaled time $\smax t$. Experimental data (points) and results of  numerical simulations \cite{Kodio2019} (solid curves) show that $\Delta A/A_0\propto \exp(2\smax t)$ at early times (dashed lines), with $\smax$ determined from \eqref{eqn:Sigma}. Here different values of the dimensionless pressure $\pring=\Tin\Rin^3/(B\t)$ are shown by different symbols: $\pring = 31$ ($\circ$), $37$ ($\square$), $56$
          ($\diamond$), $132$ ($\lhd$), $238$ ($\triangle$).  In each plot, the predicted value of $\smax(\pring)$ is used and the origin of time is shifted so that $\Delta A(\bt=0)/A_0=1\%$. (Two runs are shown for $\pring=238$, to give an indication of the systematic error.)}
  \label{fig:Area}
\end{figure}

Figure \ref{fig:Area} shows the evolution of the experimentally measured relative area change, together with numerical results from a fully nonlinear model \cite{Kodio2019} and the prediction of \eqref{eqn:Area}.  Both experiments and numerical results show the growth rate expected from \eqref{eqn:Area} at very early times and, further, that nonlinear effects (signified by significant deviations from \eqref{eqn:Area}) become important for smaller values of $\smax t$ with larger $\pring$ --- a result that is predicted by weakly nonlinear analysis \cite{Kodio2019}.

Having confirmed that the growth rate predicted from linear stability analysis corresponds to that measured experimentally, we  use numerical simulations to investigate the effect of increasing the compressive force $\Pload$ dynamically. The maximum growth rate from \eqref{eqn:GrowthRate}, $\smax=[\Pload^2/(4B\rho \t)]^{1/2}$, shows that the intrinsic time scale of instability increases with applied load, so one might expect that at early times (small load) a low mode number is excited, but  grows slowly --- before it has grown appreciably, the  compression has increased and a new fastest growing mode emerges, that grows  faster, rapidly overtaking the earlier lower mode. Since our earlier analyses have shown that ring curvature has minimal effects in the early stages of instability, we test this possibility with numerical simulations of the one-dimensional beam equation \eqref{eqn:Beam} for a beam of length $L$ with pinned boundary conditions,  $\zeta(\pm L/2,t) = \zeta_{xx}(\pm L/2,t) = 0$), and a Gaussian initial condition, subject to a time-dependent compressive force
\beq
\Pload(t)=\Pload_0 \bt^\beta,
\label{eqn:RampProfile}
\eeq with $\bt= t\times \left[B/(\rho \t)\right]^{1/2}(L/2\pi)^{-2}$ the dimensionless time.

\begin{figure}[h]
  \centering
   \includegraphics[width=0.9\linewidth]{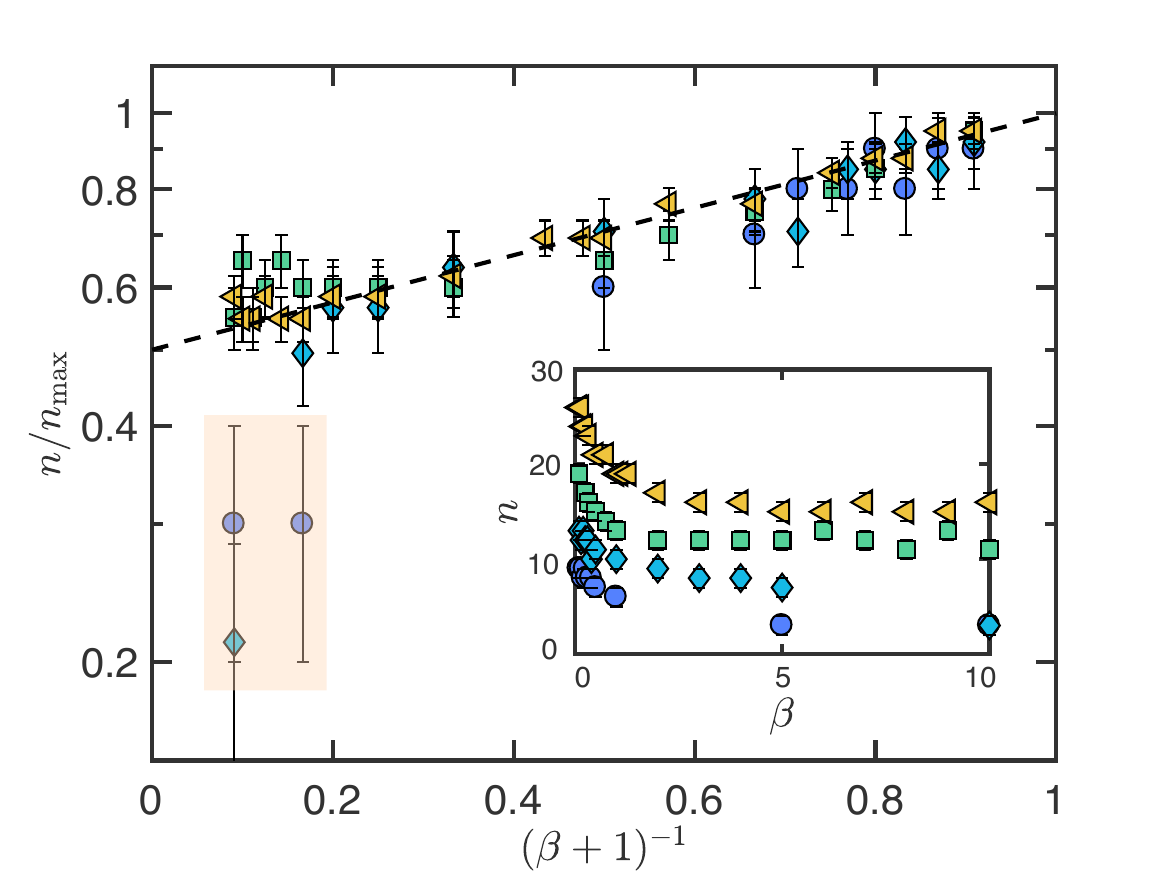}
  \caption{Mode numbers observed  with different ramping exponents $\beta$ and maximum loads $\tpinfty$. Inset: Raw data showing the numerically observed mode number at time $\bt=1$ for a loading of the form \eqref{eqn:RampProfile} with  different maximum loading pressures: $\tpinfty = 200$ ($\circ$),  $400$  ($\diamond$), $800$ ($\square$),   $1500$ ($\lhd$). Main figure: Replotting of numerical data to test the scaling prediction \eqref{eqn:nObserved}. The dashed line corresponds to  \eqref{eqn:nObserved}, with $a=0.25$ (chosen to give a reasonable description of the data). Error bars denote the observed $n\pm1$. The data highlighted in the lower left correspond to $n=3$ where the effect of the discreteness of $n$ is largest.}
  \label{fig:Ramping}
\end{figure}

The  exponent $\beta$ describes the rate of loading: $\beta\to0$ corresponds to the step-function loading assumed so far,  and $\beta=1$  to constant ramping rate.  Numerical results for the mode number observed at $\bt=1$  are shown in fig.~\ref{fig:Ramping} for a range of dimensionless maximum compressions $\tpinfty=\Pload_0L^2/(4\pi^2B)$ as a function of  $\beta$. Each value of $\tpinfty$ gives a different mode number even if imposed as a step function (see inset of fig.~\ref{fig:Ramping}); fig.~\ref{fig:Ramping} therefore shows results normalized by $\nmax=(\tpinfty/2)^{1/2}$ leading to  good collapse for large $n$. 

To understand this behaviour qualitatively, we  estimate the dimensionless time, $\btfreeze$, at which the pattern observed at $\bt=1$ is formed; we do this by requiring the argument of the exponential in our linear perturbation at $\btfreeze$,
$\sigma[\tilde{\Pload}(\btfreeze)]\cdot\btfreeze$, be comparable to that at the time of observation, $\sigma(\tpinfty)\cdot 1$, i.e.
$(\tpinfty\btfreeze^\beta) \cdot \btfreeze =a \tpinfty$ for some constant of proportionality $a$; we find $\btfreeze= a^{1/(\beta+1)}$ and, using $n=[\tilde{\Pload}(\btfreeze)/2]^{1/2}$, 
\beq
\frac{n}{\nmax}= a^{\beta/[2(\beta+1)]}.
\label{eqn:nObserved}
\eeq  We note that this scaling argument is reminiscent of the Kibble--Zurek mechanism for the  size of defects observed  during a continuous non-equilibrium phase transition as the quenching rate changes \cite{delCampo2014,Stoop2018}.

The general form of \eqref{eqn:nObserved} with $a=0.25$ is in reasonable agreement with the observations from numerical simulations (dashed line in fig.~\ref{fig:Ramping}),  although for small $n$, the errors caused by wrinkle discreteness are non-negligible.

\paragraph{Conclusion} In this Letter we have studied the dynamic buckling of elastic rings and  annuli subject to a controlled compressive force. We have shown that pattern selection in this system depends on the  geometry of the ring (its thickness-to-width ratio) as well as its mechanical properties. Although it is the presence of inertia that allows for the selection of modes of instability other than those observed in static scenarios,  the inertia itself does not affect the observed mode number when the load is applied sufficiently quickly. Otherwise, modification of the rate of loading is itself enough to select different buckling patterns --- a new approach through which different morphologies may be selected in a particular system. Moreover, because  pattern selection occurs whilst the object deforms elastically (and not plastically \cite{Lindberg1987} or fracturing brittlely \cite{Gladden2005}) changing this dynamic loading might provide a new route for changing the pattern observed in repeated runs with the same elastic object.

\begin{acknowledgments}

This research was supported by the European Research Council under the European Horizon 2020 Programme, ERC Grant Agreement no.~637334 (DV), the Leverhulme Trust via a Philip Leverhulme Prize (DV) and the Engineering and Physical Sciences Research Council grant EP/R020205/1 (AG). We are grateful to Alfonso-Castrej\'{o}n-Pita for the loan of the color high-speed camera used to take the images of fig.~2.
\end{acknowledgments}


\begin{thebibliography}{34}%
\makeatletter
\providecommand \@ifxundefined [1]{%
 \@ifx{#1\undefined}
}%
\providecommand \@ifnum [1]{%
 \ifnum #1\expandafter \@firstoftwo
 \else \expandafter \@secondoftwo
 \fi
}%
\providecommand \@ifx [1]{%
 \ifx #1\expandafter \@firstoftwo
 \else \expandafter \@secondoftwo
 \fi
}%
\providecommand \natexlab [1]{#1}%
\providecommand \enquote  [1]{``#1''}%
\providecommand \bibnamefont  [1]{#1}%
\providecommand \bibfnamefont [1]{#1}%
\providecommand \citenamefont [1]{#1}%
\providecommand \href@noop [0]{\@secondoftwo}%
\providecommand \href [0]{\begingroup \@sanitize@url \@href}%
\providecommand \@href[1]{\@@startlink{#1}\@@href}%
\providecommand \@@href[1]{\endgroup#1\@@endlink}%
\providecommand \@sanitize@url [0]{\catcode `\\12\catcode `\$12\catcode
  `\&12\catcode `\#12\catcode `\^12\catcode `\_12\catcode `\%12\relax}%
\providecommand \@@startlink[1]{}%
\providecommand \@@endlink[0]{}%
\providecommand \url  [0]{\begingroup\@sanitize@url \@url }%
\providecommand \@url [1]{\endgroup\@href {#1}{\urlprefix }}%
\providecommand \urlprefix  [0]{URL }%
\providecommand \Eprint [0]{\href }%
\providecommand \doibase [0]{http://dx.doi.org/}%
\providecommand \selectlanguage [0]{\@gobble}%
\providecommand \bibinfo  [0]{\@secondoftwo}%
\providecommand \bibfield  [0]{\@secondoftwo}%
\providecommand \translation [1]{[#1]}%
\providecommand \BibitemOpen [0]{}%
\providecommand \bibitemStop [0]{}%
\providecommand \bibitemNoStop [0]{.\EOS\space}%
\providecommand \EOS [0]{\spacefactor3000\relax}%
\providecommand \BibitemShut  [1]{\csname bibitem#1\endcsname}%
\let\auto@bib@innerbib\@empty
\bibitem [{\citenamefont {Reis}(2015)}]{Reis2015}%
  \BibitemOpen
  \bibfield  {author} {\bibinfo {author} {\bibfnamefont {P.~M.}\ \bibnamefont
  {Reis}},\ }\bibfield  {title} {\enquote {\bibinfo {title} {A perspective on
  the revival of structural (in)stability with novel opportunities for
  function: {From Buckliphobia to Buckliphilia}},}\ } {\bibfield  {journal} {\bibinfo  {journal} {J. Appl.
  Mech.}\ }\textbf {\bibinfo {volume} {82}},\ \bibinfo {pages} {111001}
  (\bibinfo {year} {2015})}\BibitemShut {NoStop}%
\bibitem [{\citenamefont {Kim}\ \emph {et~al.}(2012)\citenamefont {Kim},
  \citenamefont {Kim}, \citenamefont {Pegard}, \citenamefont {Oh},
  \citenamefont {Kagan}, \citenamefont {Fleischer}, \citenamefont {Stone},\
  and\ \citenamefont {Loo}}]{Kim2012}%
  \BibitemOpen
  \bibfield  {author} {\bibinfo {author} {\bibfnamefont {J.~B.}\ \bibnamefont
  {Kim}}, \bibinfo {author} {\bibfnamefont {P.}~\bibnamefont {Kim}}, \bibinfo
  {author} {\bibfnamefont {N.~C.}\ \bibnamefont {Pegard}}, \bibinfo {author}
  {\bibfnamefont {S.~J.}\ \bibnamefont {Oh}}, \bibinfo {author} {\bibfnamefont
  {C.~R.}\ \bibnamefont {Kagan}}, \bibinfo {author} {\bibfnamefont {J.~W.}\
  \bibnamefont {Fleischer}}, \bibinfo {author} {\bibfnamefont {H.~A.}\
  \bibnamefont {Stone}}, \ and\ \bibinfo {author} {\bibfnamefont {Y.-L.}\
  \bibnamefont {Loo}},\ }\bibfield  {title} {\enquote {\bibinfo {title}
  {{Wrinkles and deep folds as photonic structures in photovoltaics}},}\
  }\href@noop {} {\bibfield  {journal} {\bibinfo  {journal} {Nat. Photonics}\
  }\textbf {\bibinfo {volume} {6}},\ \bibinfo {pages} {327--332} (\bibinfo
  {year} {2012})}\BibitemShut {NoStop}%
\bibitem [{\citenamefont {Stafford}\ \emph {et~al.}(2004)\citenamefont
  {Stafford}, \citenamefont {Harrison}, \citenamefont {Beers}, \citenamefont
  {Karim}, \citenamefont {Amis}, \citenamefont {{van Landingham}},
  \citenamefont {Kim}, \citenamefont {Volksen}, \citenamefont {Miller},\ and\
  \citenamefont {Simonyi}}]{Stafford2004}%
  \BibitemOpen
  \bibfield  {author} {\bibinfo {author} {\bibfnamefont {C.~M.}\ \bibnamefont
  {Stafford}}, \bibinfo {author} {\bibfnamefont {C.}~\bibnamefont {Harrison}},
  \bibinfo {author} {\bibfnamefont {K.~L.}\ \bibnamefont {Beers}}, \bibinfo
  {author} {\bibfnamefont {A.}~\bibnamefont {Karim}}, \bibinfo {author}
  {\bibfnamefont {E.~J.}\ \bibnamefont {Amis}}, \bibinfo {author}
  {\bibfnamefont {M.~R.}\ \bibnamefont {{van Landingham}}}, \bibinfo {author}
  {\bibfnamefont {H.-C.}\ \bibnamefont {Kim}}, \bibinfo {author} {\bibfnamefont
  {W.}~\bibnamefont {Volksen}}, \bibinfo {author} {\bibfnamefont {R.~D.}\
  \bibnamefont {Miller}}, \ and\ \bibinfo {author} {\bibfnamefont {E.~E.}\
  \bibnamefont {Simonyi}},\ }\bibfield  {title} {\enquote {\bibinfo {title} {A
  buckling-based metrology for measuring the elastic moduli of polymeric thin
  films},}\ }\href@noop {} {\bibfield  {journal} {\bibinfo  {journal} {Nat.
  Mater.}\ }\textbf {\bibinfo {volume} {3}},\ \bibinfo {pages} {545--550}
  (\bibinfo {year} {2004})}\BibitemShut {NoStop}%
\bibitem [{\citenamefont {Huang}\ \emph {et~al.}(2007)\citenamefont {Huang},
  \citenamefont {Juskiewiscz}, \citenamefont {de~Jeu}, \citenamefont {Cerda},
  \citenamefont {Emrick}, \citenamefont {Menon},\ and\ \citenamefont
  {Russell}}]{Huang2007}%
  \BibitemOpen
  \bibfield  {author} {\bibinfo {author} {\bibfnamefont {J.}~\bibnamefont
  {Huang}}, \bibinfo {author} {\bibfnamefont {M.}~\bibnamefont {Juskiewiscz}},
  \bibinfo {author} {\bibfnamefont {W.~H.}\ \bibnamefont {de~Jeu}}, \bibinfo
  {author} {\bibfnamefont {E.}~\bibnamefont {Cerda}}, \bibinfo {author}
  {\bibfnamefont {T.}~\bibnamefont {Emrick}}, \bibinfo {author} {\bibfnamefont
  {N.}~\bibnamefont {Menon}}, \ and\ \bibinfo {author} {\bibfnamefont {T.~P.}\
  \bibnamefont {Russell}},\ }\bibfield  {title} {\enquote {\bibinfo {title}
  {Capillary wrinkling of floating thin polymer films},}\ }\href@noop {}
  {\bibfield  {journal} {\bibinfo  {journal} {Science}\ }\textbf {\bibinfo
  {volume} {317}},\ \bibinfo {pages} {650--653} (\bibinfo {year}
  {2007})}\BibitemShut {NoStop}%
\bibitem [{\citenamefont {Cerda}\ and\ \citenamefont
  {Mahadevan}(2003)}]{Cerda2003}%
  \BibitemOpen
  \bibfield  {author} {\bibinfo {author} {\bibfnamefont {E.}~\bibnamefont
  {Cerda}}\ and\ \bibinfo {author} {\bibfnamefont {L.}~\bibnamefont
  {Mahadevan}},\ }\bibfield  {title} {\enquote {\bibinfo {title} {Geometry and
  physics of wrinkling},}\ }\href@noop {} {\bibfield  {journal} {\bibinfo
  {journal} {Phys. Rev. Lett.}\ }\textbf {\bibinfo {volume} {90}},\ \bibinfo
  {pages} {074302} (\bibinfo {year} {2003})}\BibitemShut {NoStop}%
\bibitem [{\citenamefont {Davidovitch}\ \emph {et~al.}(2011)\citenamefont
  {Davidovitch}, \citenamefont {Schroll}, \citenamefont {Vella}, \citenamefont
  {Adda-Bedia},\ and\ \citenamefont {Cerda}}]{Davidovitch2011}%
  \BibitemOpen
  \bibfield  {author} {\bibinfo {author} {\bibfnamefont {B.}~\bibnamefont
  {Davidovitch}}, \bibinfo {author} {\bibfnamefont {R.~D.}\ \bibnamefont
  {Schroll}}, \bibinfo {author} {\bibfnamefont {D.}~\bibnamefont {Vella}},
  \bibinfo {author} {\bibfnamefont {M.}~\bibnamefont {Adda-Bedia}}, \ and\
  \bibinfo {author} {\bibfnamefont {E.}~\bibnamefont {Cerda}},\ }\bibfield
  {title} {\enquote {\bibinfo {title} {Prototypical model for tensional
  wrinkling in thin sheets},}\ }\href@noop {} {\bibfield  {journal} {\bibinfo
  {journal} {Proc. Natl. Acad. Sci. USA}\ }\textbf {\bibinfo {volume} {108}},\
  \bibinfo {pages} {18227--18232} (\bibinfo {year} {2011})}\BibitemShut
  {NoStop}%
\bibitem [{Note2()}]{Note2}%
  \BibitemOpen
  \bibinfo {note} {See \href{http://link.aps.org/ supplemental/10.1103/PhysRevLett.124.198003}{Supplementary
  Information}, for movies of buckling together with details of the derivation of \protect \textup
  {\hbox {\mathsurround \z@ \protect \normalfont (\ignorespaces \ref
  {eqn:annulus}\unskip \@@italiccorr )}} and illustrative numerical results
  with ramped loading \protect \textup {\hbox {\mathsurround \z@ \protect
  \normalfont (\ignorespaces \ref {eqn:RampProfile}\unskip \@@italiccorr
  )}}.}\BibitemShut {Stop}%
\bibitem [{\citenamefont {Lindberg}\ and\ \citenamefont
  {Florence}(1987)}]{Lindberg1987}%
  \BibitemOpen
  \bibfield  {author} {\bibinfo {author} {\bibfnamefont {H.~E.}\ \bibnamefont
  {Lindberg}}\ and\ \bibinfo {author} {\bibfnamefont {A.~L.}\ \bibnamefont
  {Florence}},\ }\href@noop {} {\emph {\bibinfo {title} {Dynamic Pulse
  Buckling: Theory and Experiment}}}\ (\bibinfo  {publisher} {Martinus
  Nijhoff},\ \bibinfo {address} {Dordrecht},\ \bibinfo {year}
  {1987})\BibitemShut {NoStop}%
\bibitem [{\citenamefont {Gladden}\ \emph {et~al.}(2005)\citenamefont
  {Gladden}, \citenamefont {Handzy}, \citenamefont {Belmonte},\ and\
  \citenamefont {Villermaux}}]{Gladden2005}%
  \BibitemOpen
  \bibfield  {author} {\bibinfo {author} {\bibfnamefont {J.~R.}\ \bibnamefont
  {Gladden}}, \bibinfo {author} {\bibfnamefont {N.~Z.}\ \bibnamefont {Handzy}},
  \bibinfo {author} {\bibfnamefont {A.}~\bibnamefont {Belmonte}}, \ and\
  \bibinfo {author} {\bibfnamefont {E.}~\bibnamefont {Villermaux}},\ }\bibfield
   {title} {\enquote {\bibinfo {title} {{Dynamic Buckling and Fragmentation in
  Brittle Rods}},}\ }\href@noop {} {\bibfield  {journal} {\bibinfo  {journal}
  {Phys. Rev. Lett.}\ }\textbf {\bibinfo {volume} {94}},\ \bibinfo {pages}
  {035503} (\bibinfo {year} {2005})}\BibitemShut {NoStop}%
\bibitem [{\citenamefont {Box}\ \emph {et~al.}(2013)\citenamefont {Box},
  \citenamefont {Bowman},\ and\ \citenamefont {Mullin}}]{Box2013}%
  \BibitemOpen
  \bibfield  {author} {\bibinfo {author} {\bibfnamefont {F.}~\bibnamefont
  {Box}}, \bibinfo {author} {\bibfnamefont {R.}~\bibnamefont {Bowman}}, \ and\
  \bibinfo {author} {\bibfnamefont {T.}~\bibnamefont {Mullin}},\ }\bibfield
  {title} {\enquote {\bibinfo {title} {Dynamic compression of elastic and
  plastic cellular solids},}\ }\href@noop {} {\bibfield  {journal} {\bibinfo
  {journal} {Appl. Phys. Lett.}\ }\textbf {\bibinfo {volume} {103}},\ \bibinfo
  {pages} {151909} (\bibinfo {year} {2013})}\BibitemShut {NoStop}%
\bibitem [{\citenamefont {Vermorel}\ \emph {et~al.}(2007)\citenamefont
  {Vermorel}, \citenamefont {Vandenberghe},\ and\ \citenamefont
  {Villermaux}}]{Vermorel2007}%
  \BibitemOpen
  \bibfield  {author} {\bibinfo {author} {\bibfnamefont {R.}~\bibnamefont
  {Vermorel}}, \bibinfo {author} {\bibfnamefont {N.}~\bibnamefont
  {Vandenberghe}}, \ and\ \bibinfo {author} {\bibfnamefont {E.}~\bibnamefont
  {Villermaux}},\ }\bibfield  {title} {\enquote {\bibinfo {title} {Rubber band
  recoil},}\ }\href@noop {} {\bibfield  {journal} {\bibinfo  {journal} {Proc.
  Roy. Soc. A}\ }\textbf {\bibinfo {volume} {463}},\ \bibinfo {pages}
  {641--658} (\bibinfo {year} {2007})}\BibitemShut {NoStop}%
\bibitem [{\citenamefont {Oratis}\ and\ \citenamefont
  {Bird}(2019)}]{Oratis2019}%
  \BibitemOpen
  \bibfield  {author} {\bibinfo {author} {\bibfnamefont {A.~T.}\ \bibnamefont
  {Oratis}}\ and\ \bibinfo {author} {\bibfnamefont {J.~C.}\ \bibnamefont
  {Bird}},\ }\bibfield  {title} {\enquote {\bibinfo {title} {Shooting rubber
  bands: Two self-similar retractions for a stretched elastic wedge},}\
  }\href@noop {} {\bibfield  {journal} {\bibinfo  {journal} {Phys. Rev. Lett.}\
  }\textbf {\bibinfo {volume} {122}},\ \bibinfo {pages} {014102} (\bibinfo
  {year} {2019})}\BibitemShut {NoStop}%
\bibitem [{\citenamefont {Vermorel}\ \emph {et~al.}(2009)\citenamefont
  {Vermorel}, \citenamefont {Vandenberghe},\ and\ \citenamefont
  {Villermaux}}]{Vermorel2009}%
  \BibitemOpen
  \bibfield  {author} {\bibinfo {author} {\bibfnamefont {R.}~\bibnamefont
  {Vermorel}}, \bibinfo {author} {\bibfnamefont {N.}~\bibnamefont
  {Vandenberghe}}, \ and\ \bibinfo {author} {\bibfnamefont {E.}~\bibnamefont
  {Villermaux}},\ }\bibfield  {title} {\enquote {\bibinfo {title} {Impact on
  thin elastic sheets},}\ }\href@noop {} {\bibfield  {journal} {\bibinfo
  {journal} {Proc. Roy. Soc. A}\ }\textbf {\bibinfo {volume} {465}},\ \bibinfo
  {pages} {823--842} (\bibinfo {year} {2009})}\BibitemShut {NoStop}%
\bibitem [{\citenamefont {Putelat}\ and\ \citenamefont
  {Triantafyllidis}(2014)}]{Putelat2014}%
  \BibitemOpen
  \bibfield  {author} {\bibinfo {author} {\bibfnamefont {T.}~\bibnamefont
  {Putelat}}\ and\ \bibinfo {author} {\bibfnamefont {N.}~\bibnamefont
  {Triantafyllidis}},\ }\bibfield  {title} {\enquote {\bibinfo {title} {Dynamic
  stability of externally pressurized elastic rings subjected to high rates of
  loading},}\ }\href@noop {} {\bibfield  {journal} {\bibinfo  {journal} {Int.
  J. Solids Struct.}\ }\textbf {\bibinfo {volume} {51}},\ \bibinfo {pages}
  {1--12} (\bibinfo {year} {2014})}\BibitemShut {NoStop}%
\bibitem [{Note1()}]{Note1}%
  \BibitemOpen
  \bibinfo {note} {Once the outer film is broken, the whole system is in
  free-fall for $\gtrsim 0.1\protect \mathrm {~s}$, which is longer than the
  times considered here.}\BibitemShut {Stop}%
\bibitem [{\citenamefont {Giomi}\ and\ \citenamefont
  {Mahadevan}(2012)}]{Giomi2012}%
  \BibitemOpen
  \bibfield  {author} {\bibinfo {author} {\bibfnamefont {L.}~\bibnamefont
  {Giomi}}\ and\ \bibinfo {author} {\bibfnamefont {L.}~\bibnamefont
  {Mahadevan}},\ }\bibfield  {title} {\enquote {\bibinfo {title} {Minimal
  surfaces bounded by elastic lines},}\ }\href@noop {} {\bibfield  {journal}
  {\bibinfo  {journal} {Prof. R. Soc. A}\ }\textbf {\bibinfo {volume} {468}},\
  \bibinfo {pages} {1851--1864} (\bibinfo {year} {2012})}\BibitemShut {NoStop}%
\bibitem [{\citenamefont {Chen}\ and\ \citenamefont {Fried}(2014)}]{Chen2014}%
  \BibitemOpen
  \bibfield  {author} {\bibinfo {author} {\bibfnamefont {Y.~C.}\ \bibnamefont
  {Chen}}\ and\ \bibinfo {author} {\bibfnamefont {E.}~\bibnamefont {Fried}},\
  }\bibfield  {title} {\enquote {\bibinfo {title} {Stability and bifurcation of
  a soap film spanning a flexible loop},}\ }\href@noop {} {\bibfield  {journal}
  {\bibinfo  {journal} {J. Elast.}\ }\textbf {\bibinfo {volume} {116}},\
  \bibinfo {pages} {75--100} (\bibinfo {year} {2014})}\BibitemShut {NoStop}%
\bibitem [{\citenamefont {Windenburg}\ and\ \citenamefont
  {Trilling}(1934)}]{Windenburg1934}%
  \BibitemOpen
  \bibfield  {author} {\bibinfo {author} {\bibfnamefont {D.~F.}\ \bibnamefont
  {Windenburg}}\ and\ \bibinfo {author} {\bibfnamefont {C.}~\bibnamefont
  {Trilling}},\ }\bibfield  {title} {\enquote {\bibinfo {title} {Collapse by
  instability of thin cylindrical shells under external pressure},}\
  }\href@noop {} {\bibfield  {journal} {\bibinfo  {journal} {Trans. Am. Soc.
  Mech. Eng.}\ }\textbf {\bibinfo {volume} {56}},\ \bibinfo {pages} {819--825}
  (\bibinfo {year} {1934})}\BibitemShut {NoStop}%
\bibitem [{\citenamefont {Amabili}\ and\ \citenamefont
  {Pa\"{i}doussis}(2003)}]{Amabili2003}%
  \BibitemOpen
  \bibfield  {author} {\bibinfo {author} {\bibfnamefont {M.}~\bibnamefont
  {Amabili}}\ and\ \bibinfo {author} {\bibfnamefont {M.~P.}\ \bibnamefont
  {Pa\"{i}doussis}},\ }\bibfield  {title} {\enquote {\bibinfo {title} {Review
  of studies on geometrically nonlinear vibrations and dynamics of circular
  cylindrical shells and panels, with and without fluid-structure
  interaction},}\ }\href@noop {} {\bibfield  {journal} {\bibinfo  {journal}
  {Appl. Mech. Rev.}\ }\textbf {\bibinfo {volume} {56}},\ \bibinfo {pages}
  {349--381} (\bibinfo {year} {2003})}\BibitemShut {NoStop}%
\bibitem [{\citenamefont {Carrier}(1947)}]{Carrier1947}%
  \BibitemOpen
  \bibfield  {author} {\bibinfo {author} {\bibfnamefont {G.~F.}\ \bibnamefont
  {Carrier}},\ }\bibfield  {title} {\enquote {\bibinfo {title} {On the buckling
  of elastic rings},}\ }\href@noop {} {\bibfield  {journal} {\bibinfo
  {journal} {J.  Math. Phys.}\ }\textbf {\bibinfo {volume} {26}},\ \bibinfo
  {pages} {94--103} (\bibinfo {year} {1947})}\BibitemShut {NoStop}%
\bibitem [{\citenamefont {Flaherty}\ \emph {et~al.}(1972)\citenamefont
  {Flaherty}, \citenamefont {Keller},\ and\ \citenamefont
  {Rubinow}}]{Flaherty1972}%
  \BibitemOpen
  \bibfield  {author} {\bibinfo {author} {\bibfnamefont {J.~E.}\ \bibnamefont
  {Flaherty}}, \bibinfo {author} {\bibfnamefont {J.~B.}\ \bibnamefont
  {Keller}}, \ and\ \bibinfo {author} {\bibfnamefont {S.~I.}\ \bibnamefont
  {Rubinow}},\ }\bibfield  {title} {\enquote {\bibinfo {title} {Post buckling
  behavior of elastic tubes and rings with opposite sides in contact},}\
  }\href@noop {} {\bibfield  {journal} {\bibinfo  {journal} {SIAM J. Appl.
  Math.}\ }\textbf {\bibinfo {volume} {23}},\ \bibinfo {pages} {446--455}
  (\bibinfo {year} {1972})}\BibitemShut {NoStop}%
\bibitem [{\citenamefont {Hazel}\ and\ \citenamefont
  {Mullin}(2017)}]{Hazel2017}%
  \BibitemOpen
  \bibfield  {author} {\bibinfo {author} {\bibfnamefont {A.~L.}\ \bibnamefont
  {Hazel}}\ and\ \bibinfo {author} {\bibfnamefont {T.}~\bibnamefont {Mullin}},\
  }\bibfield  {title} {\enquote {\bibinfo {title} {On the buckling of elastic
  rings by external confinement},}\ }\href@noop {} {\bibfield  {journal}
  {\bibinfo  {journal} {Phil. Trans. R. Soc.}\ }\textbf {\bibinfo {volume}
  {375}},\ \bibinfo {pages} {20160227} (\bibinfo {year} {2017})}\BibitemShut
  {NoStop}%
\bibitem [{\citenamefont {Pi{\~{n}}eirua}\ \emph {et~al.}(2013)\citenamefont
  {Pi{\~{n}}eirua}, \citenamefont {Tanaka}, \citenamefont {Roman},\ and\
  \citenamefont {Bico}}]{Pineirua2013}%
  \BibitemOpen
  \bibfield  {author} {\bibinfo {author} {\bibfnamefont {M.}~\bibnamefont
  {Pi{\~{n}}eirua}}, \bibinfo {author} {\bibfnamefont {N.}~\bibnamefont
  {Tanaka}}, \bibinfo {author} {\bibfnamefont {B.}~\bibnamefont {Roman}}, \
  and\ \bibinfo {author} {\bibfnamefont {J.}~\bibnamefont {Bico}},\ }\bibfield
  {title} {\enquote {\bibinfo {title} {{Capillary buckling of a floating
  annulus}},}\ } {\bibfield  {journal}
  {\bibinfo  {journal} {Soft Matter}\ }\textbf {\bibinfo {volume} {9}},\
  \bibinfo {pages} {10985} (\bibinfo {year} {2013})}\BibitemShut {NoStop}%
\bibitem [{\citenamefont {Paulsen}\ \emph {et~al.}(2017)\citenamefont
  {Paulsen}, \citenamefont {D\'{e}mery}, \citenamefont {Toga}, \citenamefont
  {Qiu}, \citenamefont {Russell}, \citenamefont {Davidovitch},\ and\
  \citenamefont {Menon}}]{Paulsen2017}%
  \BibitemOpen
  \bibfield  {author} {\bibinfo {author} {\bibfnamefont {J.~D.}\ \bibnamefont
  {Paulsen}}, \bibinfo {author} {\bibfnamefont {V.}~\bibnamefont {D\'{e}mery}},
  \bibinfo {author} {\bibfnamefont {K.~B.}\ \bibnamefont {Toga}}, \bibinfo
  {author} {\bibfnamefont {Z.}~\bibnamefont {Qiu}}, \bibinfo {author}
  {\bibfnamefont {T.~P.}\ \bibnamefont {Russell}}, \bibinfo {author}
  {\bibfnamefont {B.}~\bibnamefont {Davidovitch}}, \ and\ \bibinfo {author}
  {\bibfnamefont {N.}~\bibnamefont {Menon}},\ }\bibfield  {title} {\enquote
  {\bibinfo {title} {Geometry-driven folding of a floating annular sheet},}\
  }\href@noop {} {\bibfield  {journal} {\bibinfo  {journal} {Phys. Rev. Lett.}\
  }\textbf {\bibinfo {volume} {118}},\ \bibinfo {pages} {048004} (\bibinfo
  {year} {2017})}\BibitemShut {NoStop}%
\bibitem [{\citenamefont {Howell}\ \emph {et~al.}(2009)\citenamefont {Howell},
  \citenamefont {Kozyreff},\ and\ \citenamefont {Ockendon}}]{Howell2009}%
  \BibitemOpen
  \bibfield  {author} {\bibinfo {author} {\bibfnamefont {Peter}\ \bibnamefont
  {Howell}}, \bibinfo {author} {\bibfnamefont {Gregory}\ \bibnamefont
  {Kozyreff}}, \ and\ \bibinfo {author} {\bibfnamefont {John}\ \bibnamefont
  {Ockendon}},\ }\href@noop {} {\emph {\bibinfo {title} {Applied Solid
  Mechanics}}}\ (\bibinfo  {publisher} {Cambridge University Press},\ \bibinfo
  {year} {2009})\BibitemShut {NoStop}%
\bibitem [{Note3()}]{Note3}%
  \BibitemOpen
  \bibinfo {note} {For large $\protect \mathaccentV {tilde}07E{p}_{\protect
  \mathrm {ring}}$, coarsening of the buckles in one portion of the ring
  occurred before the outer soap film had fully retracted. In such cases, $n$
  was inferred from the measured buckling wavelength at onset; in other cases,
  the value of $n$ reported is the maximum number of buckles
  observed.}\BibitemShut {Stop}%
\bibitem [{\citenamefont {Kantsler}\ and\ \citenamefont
  {Goldstein}(2012)}]{Kantsler2012}%
  \BibitemOpen
  \bibfield  {author} {\bibinfo {author} {\bibfnamefont {V.}~\bibnamefont
  {Kantsler}}\ and\ \bibinfo {author} {\bibfnamefont {R.~E.}\ \bibnamefont
  {Goldstein}},\ }\bibfield  {title} {\enquote {\bibinfo {title} {Fluctuations,
  dynamics, and the stretch-coil transition of single actin filaments in
  extensional flows},}\ }\href@noop {} {\bibfield  {journal} {\bibinfo
  {journal} {Phys. Rev. Lett.}\ }\textbf {\bibinfo {volume} {108}},\ \bibinfo
  {pages} {038103} (\bibinfo {year} {2012})}\BibitemShut {NoStop}%
\bibitem [{\citenamefont {Hutchinson}\ and\ \citenamefont
  {Budiansky}(1966)}]{Hutchinson1966}%
  \BibitemOpen
  \bibfield  {author} {\bibinfo {author} {\bibfnamefont {J.~W.}\ \bibnamefont
  {Hutchinson}}\ and\ \bibinfo {author} {\bibfnamefont {B.}~\bibnamefont
  {Budiansky}},\ }\bibfield  {title} {\enquote {\bibinfo {title} {Dynamic
  buckling estimates},}\ }\href@noop {} {\bibfield  {journal} {\bibinfo
  {journal} {AIAA J.}\ }\textbf {\bibinfo {volume} {4}},\ \bibinfo {pages}
  {525--530} (\bibinfo {year} {1966})}\BibitemShut {NoStop}%
\bibitem [{\citenamefont {Karagiozova}\ and\ \citenamefont
  {Alves}(2008)}]{Karagiozova2008}%
  \BibitemOpen
  \bibfield  {author} {\bibinfo {author} {\bibfnamefont {D.}~\bibnamefont
  {Karagiozova}}\ and\ \bibinfo {author} {\bibfnamefont {M.}~\bibnamefont
  {Alves}},\ }\bibfield  {title} {\enquote {\bibinfo {title} {Dynamic
  elastic-plastic buckling of structural elements: A review},}\ }\href@noop {}
  {\bibfield  {journal} {\bibinfo  {journal} {Appl. Mech. Rev.}\ }\textbf
  {\bibinfo {volume} {61}},\ \bibinfo {pages} {040803} (\bibinfo {year}
  {2008})}\BibitemShut {NoStop}%
\bibitem [{\citenamefont {Wah}(1970)}]{Wah1970}%
  \BibitemOpen
  \bibfield  {author} {\bibinfo {author} {\bibfnamefont {T.}~\bibnamefont
  {Wah}},\ }\bibfield  {title} {\enquote {\bibinfo {title} {Dynamic buckling of
  thin circular rings},}\ }\href@noop {} {\bibfield  {journal} {\bibinfo
  {journal} {Int. J. Mech. Sci.}\ }\textbf {\bibinfo {volume} {12}},\ \bibinfo
  {pages} {143--155} (\bibinfo {year} {1970})}\BibitemShut {NoStop}%
\bibitem [{\citenamefont {Boys}(1958)}]{Boys1958}%
  \BibitemOpen
  \bibfield  {author} {\bibinfo {author} {\bibfnamefont {C.~V.}\ \bibnamefont
  {Boys}},\ }\href@noop {} {\emph {\bibinfo {title} {Soap Bubbles: Their
  colours and forces which mold them}}}\ (\bibinfo  {publisher} {Dover},\
  \bibinfo {address} {Mineola, New York},\ \bibinfo {year} {1958})\BibitemShut
  {NoStop}%
\bibitem [{\citenamefont {Kodio}\ \emph {et~al.}(2019)\citenamefont {Kodio},
  \citenamefont {Goriely},\ and\ \citenamefont {Vella}}]{Kodio2019}%
  \BibitemOpen
  \bibfield  {author} {\bibinfo {author} {\bibfnamefont {O.}~\bibnamefont
  {Kodio}}, \bibinfo {author} {\bibfnamefont {A.}~\bibnamefont {Goriely}}, \
  and\ \bibinfo {author} {\bibfnamefont {D.}~\bibnamefont {Vella}},\ }\bibfield
   {title} {\enquote {\bibinfo {title} {Dynamic buckling of an inextensible elastic ring: Linear and nonlinear analyses},}\
  }\href@noop{} {\bibfield  {journal} {\bibinfo  {journal} {Phys. Rev. E}\
  }\textbf {\bibinfo {volume} {101}},\ \bibinfo {pages} {053002} (\bibinfo
  {year} {2020})} \BibitemShut {NoStop}%
\bibitem [{\citenamefont {{del Campo}}\ and\ \citenamefont
  {Zurek}(2014)}]{delCampo2014}%
  \BibitemOpen
  \bibfield  {author} {\bibinfo {author} {\bibfnamefont {A.}~\bibnamefont {{del
  Campo}}}\ and\ \bibinfo {author} {\bibfnamefont {W.~H.}\ \bibnamefont
  {Zurek}},\ }\bibfield  {title} {\enquote {\bibinfo {title} {Universality of
  phase transition dynamics: Topological defects from symmetry breaking},}\
  }\href@noop {} {\bibfield  {journal} {\bibinfo  {journal} {Int. J. Mod.
  Phys.}\ }\textbf {\bibinfo {volume} {29}},\ \bibinfo {pages} {1430018}
  (\bibinfo {year} {2014})}\BibitemShut {NoStop}%
\bibitem [{\citenamefont {Stoop}\ and\ \citenamefont
  {Dunkel}(2018)}]{Stoop2018}%
  \BibitemOpen
  \bibfield  {author} {\bibinfo {author} {\bibfnamefont {N.}~\bibnamefont
  {Stoop}}\ and\ \bibinfo {author} {\bibfnamefont {J.}~\bibnamefont {Dunkel}},\
  }\bibfield  {title} {\enquote {\bibinfo {title} {Defect formation dynamics in
  curved elastic surface crystals},}\ }\href@noop {} {\bibfield  {journal}
  {\bibinfo  {journal} {Soft Matter}\ }\textbf {\bibinfo {volume} {14}},\
  \bibinfo {pages} {2329--2338} (\bibinfo {year} {2018})}\BibitemShut {NoStop}%
\end{thebibliography}

%

\end{document}